\documentstyle[12pt,epsfig]{article}
\newcommand{\be}{\begin{equation}}
\newcommand{\ee}{\end{equation}}
\newcommand{\ba}{\begin{eqnarray}}
\newcommand{\ea}{\end{eqnarray}}
\newcommand{\no}{\nonumber \\}
\renewcommand{\thefootnote}{\fnsymbol{footnote}}
\begin{document}
\begin{titlepage}
\pagestyle{empty}
\vspace{1.0in}
\begin{flushright}
\today
\end{flushright}
\vspace{1.0in}
\begin{center}
\begin{large}
{\bf{A SCHEMATIC MODEL FOR DENSITY}}\\
 {\bf DEPENDENT VECTOR MESON MASSES}\footnote{Talk given by G.E. Brown at
the AIP/KKG Memorial Meeting, 3 October 1998, who dedicated this work
to Klaus Kinder--Geiger, long-time friend and stimulating colleague.}\\
\end{large}
\vskip 1cm
Y. Kim$^{(a,b)}$, R. Rapp$^{(b)}$, G.E. Brown$^{(b)}$ and Mannque Rho$^{(c)}$
\vskip 0.2cm
{\it (a)  Department of Physics, Hanyang University, Seoul 133-791, Korea}

{\it (b) Department of Physics and Astronomy, State University of New York\\
Stony Brook, NY 11794-3800, USA}

{\it (c) Service de Physique Th\'eorique, CE Saclay\\
91191 Gif-sur-Yvette, France}
\end{center}
\vskip 2cm

\centerline{\bf Abstract}
\vskip 0.1cm

\noindent A schematic two-level model consisting of a ``collective" bosonic
state and an ``elementary" meson is constructed that provides
interpolation from a hadronic description (a la Rapp/Wambach) to B/R scaling
for the description of
properties of vector mesons in dense medium. The development is based on
a close analogy to the degenerate schematic model of Brown for giant resonances
in nuclei.

\end{titlepage}
\setcounter{footnote}{0}
\renewcommand{\thefootnote}{\arabic{footnote}}
\section{Introduction}

\indent\indent
The density dependence of vector-meson masses suggested by
Brown/Rho (B/R) scaling \cite{br} stimulated a lot of interest. In
particular, the CERES dilepton experiments \cite{ceres} provided
strong evidence that the properties of the $\rho$ mesons are
nontrivially modified in hadronic matter. An excess of the dilepton
with low invariant mass, as well as strength missing from the
region of the free $\rho$-mass, are found in the experiments,
although these determinations are not very quantitative up to now.
However, experiments now underway with TPC should determine with
good accuracy just how much strength is left at the free
$\rho$-meson pole during the time of overlap of the heavy nuclei up
until freezeout, that is, during the fireball.

The simplest and most economical explanation for the observed
low-mass dileptons is given in terms of quasiparticles (both
fermions and bosons) whose masses drop according to B/R scaling,
thereby making an appealing link to the chiral structure of the
hadronic vacuum. In an alternative view to this description, Rapp,
Chanfray and Wambach (R/W)~\cite{rapp} claimed that the excess of
low-mass dileptons can also follow from conventional many-body
physics. On a rather general ground, this ``alternative''
description was in a sense anticipated as discussed by one of the
authors~\cite{duality}. In analogy to the quark-hadron duality in
heavy-light meson decay processes, one may view B/R scaling as a
``partonic'' picture while R/W as a hadronic one. One way of
succinctly summarizing the situation is that the former is a
top-down approach and the latter a bottom-up one. The link between
B/R scaling and the Landau quasiparticle interaction $F_1$
established in \cite{FR96} is one specific indication for this
``duality.'' Indeed, in \cite{brown}, Brown et al argued that the
R/W explanation could be interpreted as a density-dependent
$\rho$-meson mass, calculated in a hadron language (in contrast to
that of constituent quarks used by Brown and Rho). In particular it
was suggested in ref.\cite{brown} that if one replaced the
$\rho$-meson mass $m_\rho$ by the mass $m_\rho^*(\rho)$ at the
density being considered, one would arrive at a description, in
hadron language, which at high densities appeared dual to that of
the Brown/Rho one in terms of constituent quarks. These
developments involved the interpretation of a {\it collective}
isobar-hole excitation as an effective vector meson field operating
on the ground state of the nucleus; {\it i.e.},
\be
\frac{1}{\sqrt{A}}\sum_{i}[N^*(1520)_iN_i^{-1}]^{1-}
\simeq \sum_{i} [\rho(x_i)~ or~ \omega(x_i) ]|\Psi_0>_s,\label{col}
\ee
with the antisymmetrical (symmetrical) sum over neutrons and
protons giving the $\rho$-like ($\omega$-like) nuclear excitation.
The dropping vector meson masses could then be calculated in terms
of mixing of the nuclear collective state, eq.(\ref{col}), with the
elementary vector meson through the mixing matrix elements of fig.
1.
\begin{figure}
\centerline{\epsfig{file=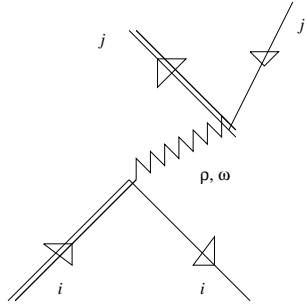,width=4cm}}
\caption{\small The mixing matrix element $M_{ij}$}\label{coupling}
\end{figure}
Now building up the collective nuclear mode, the latter can be
identified as an analog to the state in the degenerate schematic
model of Brown for giant dipole resonance~\cite{gerry}. An
important development which leads to the assumption eq.(\ref{col})
was furnished by Friman, Lutz and Wolf\cite{friman}. From empirical
values of the amplitudes such as $\pi + N\rightarrow \rho + N$,
etc. they constructed the $\rho$-like or $\omega$-like states
encouraged by our assumption eq.(\ref{col}). Thus our input
assumption receives substantial empirical support. Furthermore, one
can obtain the coupling constants of the nucleon to three
collective states from their work, that to the $\rho$-like
excitation being close to the one used by Brown et al in
\cite{brown}.

In this paper, we reformulate the heuristic idea described in
\cite{brown} in a more specific form by clearly stating the set
of assumptions we make in implementing the strategy. Our principal
aim is to construct a model that interpolates the R/W theory valid
near zero density to the B/R theory valid near the chiral phase
transition density. Within the schematic two-level (``collective"
field and ``elementary" field) model defined by the coupling matrix
element $M_{ij}$ of fig.\ref{coupling}, we assume that the
self-energy $\Sigma$, eq.(\ref{RWself}), that enters the dispersion
formula (to be given below) encodes the mechanism to interpolate
between the two regimes. In \cite{brown}, it was suggested~\footnote{
We have no convincing argument for the validity of this procedure. Our 
conjecture is as follows. To zeroth order in density, the $\rho N^* N$ 
coupling is of the form $\frac{f}{m}q_0$ with a dimensionless constant
$f$ and $q_0$ is the fourth component of the four-vector of the $\rho$ meson.
If one writes this as $Fq_0$ with $F=f/m$, then one should compute
the medium renormalization of the constant $F$ which will then depend on
density $\rho$. In order for the vector meson mass to go to zero at some 
high density so as to match B/R scaling, 
it is required that $F(\rho)q_0 \rightarrow {\rm constant}\neq
0$. For $q=|\vec{q}|\approx 0$ which we are considering, this can be
satisfied if $F(\rho)\sim {m^*}^{-1}$, modulo an overall constant. This is
essentially the essence of the proposal of ref.\cite{brown}}, that
going to B/R scaling from R/W theory corresponds to replacing the
$m_\rho^2$ appearing in the denominator of (\ref{RWself}) by
${m_\rho^*}^2$. In this paper, we shall show that this is indeed
consistent with what is expected at $\rho\sim 0$ and $\rho\sim
\rho_c$. Specifically, with the present construction, $m_\rho^*$
goes to zero at $\rho_c\sim 2.75\rho_0$ as in the
Nambu-Jona-Lasinio calculation~\cite{rho}. Without this replacement
in R/W, however, the $\rho$ mass can never go to zero at any
density. This is because the self-energy is prefixed by
$q_0^2\rightarrow {m_\rho^*}^2$, so that it would vanish if
$m_\rho^*\rightarrow 0$, resulting in a contradiction. Furthermore
at any density, there will always be two states of the $\rho$
quantum number in R/W whereas in B/R, all of the strength
($A(\omega)$ defined later) goes into the lower one as
$m_\rho^*\rightarrow 0$ with the width going to zero as well since
the phase space for decay goes to zero. We identify this state as
the {\it effective} $\rho$ degree of freedom as one approaches the
critical density. At lower densities, we cannot make this
identification, because of the two different $\rho$-states, so our
model has a clear interpretation only at $\rho\approx 0$ and
$\rho\approx \rho_c$.

Since we have a simple schematic model which can describe (roughly)
the Rapp/Wambach or Brown/Rho regimes, depending on whether one
scales with $m_\rho$ or $m_\rho^*$, we can easily calculate the
general amount of strength to be found in low-mass dileptons, and
the strength removed from the free-$\rho$ pole. We adopt the
following strategy.

We calculate the weighting factor $Z$ for the two states,
nuclear collective and elementary vector, which mix.
The large imaginary part of the energy of the former
state makes it difficult to show in detail exactly
how its strength is distributed, but we know that
the amount of the strength in that state -- in our
two-level model --  must be just the strength removed from
the higher state by the mixing. This strength will be
formed at low invariant masses. We make rough estimates, by
including or not including various widths, of the energies
at which this lower strength will be formed.

We note here that in our two-level model, the state originating
from the elementary $\rho$ (or $\omega$) is pushed up substantially
in energy. We believe much of this displacement to be an artefact
of our two-level model, because there is substantial strength with
$\rho$ (or $\omega$) quantum numbers lying above the single $\rho$
(or $\omega$) excitation we have chosen and the strength above will
push down the upper $\rho$ (or $\omega$). Whereas some shift upwards of the
strength originally in the elementary $\rho$ (or $\omega$) may be
formed, our two-level model certainly will overdo the shift. We do
not believe this defect to greatly change the amount of strength
shifted to lower invariant mass, however.

Of course the total strength is conserved, so the amount of
strength shifted to lower energies must be that missing from the
higher state. However we note that the spectral strength
$A(\omega)$ is related to $Z(\omega)$ by a factor
\be
A(\omega)=\frac{Z(\omega)}{2\omega}.
\ee
This is clear because the sum rule on the $A(\omega)$, essentially
oscillator strength, must be just that of the (energy weighted)
Thomas-Reiche- Kuhn sum rule. It is the quantity $A(\omega)$ which
enters into the rate equation for the dilepton production. Thus if
the nuclear collective state is pushed down to an energy
$\omega\simeq m_\rho/2$ with $25 \%$ of the strength being removed
from the elementary $\rho$ pole, then one finds roughly equal
spectral weights in the low-energy region and in the region of the
elementary $\rho$. Because of the much larger Boltzmann factor in
the low-energy region, a factor of several more dileptons will come
from it than from the $\rho$-pole, given temperature $T\sim 150$
MeV.

\section {The $\rho$-Meson in Nuclear Matter}
\indent\indent
The in-medium $\rho$-meson propagator is given by,
\be
D_\rho(q_0,\vec q)=1/[q_0^2-\vec q^2
-(m_\rho^0)^{2}-\Sigma_{\pi\pi}(q_0,\vec q)-\Sigma_{\rho N^* N}(q_0,\vec q)
]
\ee
where $m_\rho^0$ is a bare mass. The real part of $\Sigma_{\pi\pi}$
is taken into account approximately by defining
$m_\rho^2=(m_\rho^0)^2+{\rm Re}\Sigma_{\pi\pi}=(770)^2\ {\rm
MeV}^2$ \cite{peters}. The imaginary part is taken to be
\be
{\rm Im}\Sigma_{\pi\pi}(q_0,\vec q)=-m_\rho\Gamma_{\pi\pi}(q_0,\vec
q).
\ee
Then we get,\footnote{From the particle data book, we find
$\Gamma_{\pi\pi} =150$ MeV (full width). In ref.\cite{rapp}
(compare this with that in ref.\cite{peters} ), the authors used
the following form for $\Gamma_{\pi\pi}$
\be
\Gamma_{\pi\pi}=\frac{p(q_0)^3}{p_0^3} (\frac{2\Lambda_\rho^2+m_\rho^2}
{2\Lambda_\rho^2+q_0^2})\Gamma_{\pi\pi}^0
\ee
with $\Gamma_{\pi\pi}^0=120$ MeV and $p_0\equiv p(q_0=m_\rho)$.
Here $p$ refers to the pion momentum ($p=|\vec p|$) and $q_0$
the energy of the $\rho$-meson.}
\be
D_\rho(q_0,\vec q)=1/[q_0^2-\vec q^2 -m_\rho^{2}(q_0)+im_\rho
\Gamma_{\pi\pi}(q_0)-\Sigma_{\rho N^* N}]\label{rhop1}
\ee
where $m_\rho (q_0)$ is the energy-dependent mass with the energy
dependence lodged in the self-energy. The $\rho$-meson dispersion
relation( at $\vec q =0$) is given by
\ba
q_0^2 =m_\rho^2+{\rm Re} \Sigma_{\rho N^* N}(q_0).
\ea
Solving this equation is equivalent to determining the zeros in the
real part of the inverse $\rho$-meson propagator, eq.(\ref{rhop1}).

\subsection{The Rapp/Wambach approach}
\indent\indent
We start with the crucial ingredients in R/W (Rapp/Wambach) theory
\cite{brown}. The $\rho$-meson self-energy coming from the
particle-hole excitation $N^*(1520)N^{-1}$ is
\be
\Sigma_{\rho N^* N}(q_0)=f_{\rho N^* N}^2\frac{8}{3}
\frac{q_0^2}{m_\rho^2}\frac{\rho_0}{4}
(\frac{2(\Delta E)}{(q_0+i\Gamma_{tot}/2)^2-(\Delta
E)^2})\label{RWself}
\ee
where $\Delta E=M_{N^*}-M_N\simeq 1520-940=580$ MeV and
$\Gamma_{tot}=\Gamma_0 +\Gamma_{med}$ where $\Gamma_0$ is the full
width of $N^*(1520)$ in free space, $\sim 120$ MeV. The
$\Gamma_{med}$ represents medium corrections to the width of $N^*(1520)$
\cite{peters}. In this calculation, we shall just replace
integration over fermi-momentum by nuclear density $\rho_0$, an
approximation presumably good at low density. The real part of
$\Sigma_{\rho N^* N}(q_0)$ then takes the form
\def\Re{\rm Re}\def\Im{\rm Im}
\be
\Re\Sigma_{\rho N^* N}=f_{\rho N^* N}^2\frac{4}{3}\frac{q_0^2}{m_\rho^2}\rho_0
\frac{\Delta E(q_0^2-(\Delta E)^2-\frac{1}{4}\Gamma_{tot}^2)}
{(q_0^2-(\Delta E)^2-\frac{1}{4}\Gamma_{tot}^2)^2+\Gamma_{tot}^2q_0^2}
\ee
This leads to the $\rho$-meson dispersion relation (for $\vec q
=0$)
\ba
q_0^2 &=&m_\rho^2+\Re \Sigma_{\rho N^* N}(q_0)\no
&=&m_\rho^2+f_{\rho N^* N}^2\frac{4}{3}\frac{q_0^2}{m_\rho^2}\rho_0
\frac{\Delta E(q_0^2-(\Delta E)^2-\frac{1}{4}\Gamma_{tot}^2)}
{(q_0^2-(\Delta
E)^2-\frac{1}{4}\Gamma_{tot}^2)^2+\Gamma_{tot}^2q_0^2}.\label{dis}
\ea

The $Z$-factor that represents the spectral weight of the upper state is,
in general, defined by
\be
Z=(1-\frac{\partial\Sigma}{\partial q_0^2})^{-1} .
\ee
To get this quantity, we first evaluate $\frac{\partial\Sigma_{\rho
N^* N}}{\partial q_0^2}$. Due to the width of $N^*(1520)$,
$\Sigma_{\rho N^* N}$ has an imaginary part. We shall define
\be
Z=(1-\frac{\partial}{\partial q_0^2}\Re \Sigma_{\rho N^* N})^{-1}
\ee
taking the real part of $\Sigma_{\rho N^* N}$ so as to make the
$Z$-factor real
\footnote{There is a point
which should be clarified. If we solve the equation of the
$\rho$-meson dispersion relation (eq.(\ref{dis})), we will possibly
get one real and two complex valued solutions for $q_0$. For
the real solution, our definition of $Z$-factor (eq.(\ref{zz}))
could be correct. But in the case of the complex solution,  we are
not sure whether this definition still makes sense. Of course, we
can use a sum rule for $Z$-factor to estimate the $Z$-factor
corresponding to complex solutions. This point needs further
study.}. Defining $x=\frac{q_0^2}{m_\rho^2}$, we get
\ba
\frac{\partial}{\partial q_0^2} \Re\Sigma_{\rho N^* N}
&=&\frac{\partial}{\partial x}(\frac{\Re \Sigma_{\rho N^*
N}}{m_\rho^2})\no &=& c\frac{(2x-c_1
-c_2)((x-c_1-c_2)^2+4c_2x)}{((x-c_1-c_2)^2+4c_2x)^2}\no &
&-\frac{x(x-c_1-c_2)
(2(x-c_1-c_2)+4c_2)}{((x-c_1-c_2)^2+4c_2x)^2}\label{zz}
\ea
where $c=f_{\rho N^* N}^2\frac{4}{3}\frac{\rho_0}{m_\rho^3}
\frac{\Delta E}{m_\rho}$, $c_1=\frac{(\Delta E)^2}{m_\rho^2}\simeq 0.567$
 and $c_2=\frac{1}{4}\frac{\Gamma_{tot}^2}{m_\rho^2}$.
We can readily obtain the zeros in the real part of the
$\rho$-propagator and the $Z$ factor by plotting figures like those
of Fig.3 in ref.\cite{brown}. But here we shall get them by
directly solving eq.(\ref{dis}) and calculating eq.(\ref{zz}).
\vskip0.3cm

$\bullet$ \underline{Case of $\Gamma_{tot}=0$}
\vskip 0.3cm

For simplicity, let us set $\Gamma_{tot}=0$. The relevant equations
simplify to
\ba
q_0^2&=&m_\rho^2+f_{\rho N^*N}^2\frac{4}{3}\frac{q_0^2}{m_\rho^2}
\rho_0\frac{\Delta E}{q_0^2 -(\Delta E)^2},\no
\frac{\partial\Sigma_{\rho N^*N}}{\partial q_0^2}&=&
\frac{\partial}{\partial x}(\frac{\Sigma_{\rho N^* N}}{m_\rho^2})\no
&=&f_{\rho N^* N}^2\frac{4}{3}\frac{\rho_0^2}{m_\rho^3}\frac{\Delta E}{m_\rho}
\frac{-\frac{(\Delta E)^2}{m_\rho^2}}{(x-\frac{(\Delta E)^2}{m_\rho^2})^2}.
\ea
Written in terms of the quantity $x\equiv q_0^2/m_\rho^2$, the
dispersion relation reads
\be
x=1+0.208x\frac{1}{x-0.567}
\ee
where we have used $\frac{f_{\rho N^*N}^2}{4\pi}=5.5$ from
ref.\cite{brown} and $\rho_0 \simeq
\frac{1}{2}m_\pi^3$. The solutions are
\be
q_0^-\simeq 498~{\rm MeV}, ~q_0^+\simeq 897~{\rm MeV}.
\ee
The formula for $Z$-factor
\be
Z=(1+0.118\frac{1}{(x-0.567)^2})^{-1}
\ee
yields the corresponding $Z$-factors
\be
Z(q_0^-)\simeq 0.16,~Z(q_0^+)\simeq 0.84.
\ee
Naively extrapolated to a higher density, say, $\rho\simeq
2.5\rho_0$, the results come out to be
\be
q_0^-\simeq 436\ {\rm MeV}, ~q_0^+\simeq 1023\ {\rm MeV}
\ee
and
\be
Z(q_0^-)\simeq 0.17,~Z(q_0^+)\simeq 0.83
\ee
\vskip 0.3cm

{$\bullet$} \underline{Case of $\Gamma_{tot}=\Gamma_0=120$ MeV}
\vskip 0.3cm

Substituting $\Gamma_0=120$ MeV and $\Gamma_{med}=0$ into the
dispersion relation at normal nuclear density $\rho_0$
\be
x=1+0.208x\frac{x-0.567-\frac{\Gamma_{tot}^2}{4m_\rho^2}}
{(x-0.567-\frac{\Gamma_{tot}^2}{4m_\rho^2})^2+\frac{\Gamma_{tot}^2}{m_\rho^2}x}
\ee
we obtain the solutions
\be
x=\frac{q_0^2}{m_\rho^2}= 1.24,~0.49+ i0.267,~0.49- i0.267
\ee
where $i$ refers to imaginary. Taking the real part of the
solutions, we get\footnote{As stated, we do not know how to
interpret physically the imaginary parts of the solution. Of
course, the imaginary part tells us that the pole is located off
the real axis.}
\be
q_0^-\simeq \sqrt{0.49*m_\rho^2}=541\ {\rm MeV}, ~q_0^+\simeq 857\
{\rm MeV}.
\ee

\begin{figure}[t]
\centerline{\epsfig{file=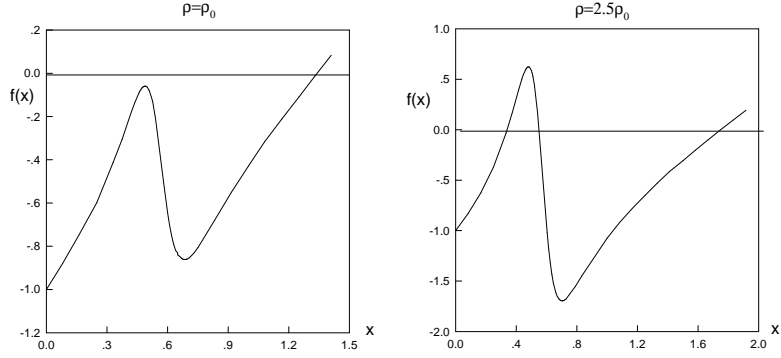,width=11cm}}
\vskip -1.4cm
\caption{\small $\rho$-meson dispersion relation for $\Gamma_0=120$ MeV
 at $\rho=\rho_0$ and $\rho=2.5\rho_0$.
 The horizontal axis $x$ represents $\frac{q_0^2}{m_\rho^2}$ and the
vertical axis $f(x)=x-1-\Re\Sigma_{\rho N^*N}(x)$.}\label{fig23}
\end{figure}

The $Z$-factor for $q_0^+$ state is calculated to be
\be
Z(q_0^+)\simeq 0.86
\ee
with the remaining strength going to the lower state.  For
$\rho=2.5\rho_0$, we get
\be
x=0.34,~0.55,~1.7
\ee
corresponding to
\be
q_0=489\ {\rm MeV}, ~571\ {\rm MeV}, ~1003\ {\rm MeV}.
\ee
The corresponding $Z$-factors are\footnote{We interpret these three
states of $\rho$-meson quantum number to be the ``elementary" $\rho
,~N^*(1520)N^{-1}\pi N$ and $N^*(1520)N^{-1}$. See fig.\ref{fig23}}
\ba
Z(449)&=&0.21\no Z(571)&=&-0.056\no Z(1004)&=&0.83.
\ea
The $\rho$-meson dispersion relation with $\Gamma_0=120MeV$ at
$\rho=\rho_0$ and $\rho=2.5\rho_0$ is shown in fig.\ref{fig23}.

\subsection{The B/R approach}
\indent\indent
As stated in Introduction, we propose that approaching B/R-scaling
from hadronic excitations is effected by replacing
$\frac{q_0^2}{m_\rho^2}$ by $1$ in the $\Sigma_{\rho N^* N}$ that
enters in the dispersion relation \cite{brown}. Let us see how this
ansatz works out in reproducing the structure of B/R scaling as
density increases. From (\ref{dis}), we get a minimally modified
dispersion relation for the $\rho$-meson in medium
\be
q_0^2
=m_\rho^2+f_{\rho N^* N}^2\frac{4}{3}\rho_0
\frac{\Delta E(q_0^2-(\Delta E)^2-\frac{1}{4}\Gamma_{tot}^2)}
{(q_0^2-(\Delta
E)^2-\frac{1}{4}\Gamma_{tot}^2)^2+\Gamma_{tot}^2q_0^2}\label{dis1}.
\ee
\vskip 0.3cm
In this formula, we shall assume that in medium, $\Delta E$ remains unchanged
(assumption valid to the leading order in $1/N_c$) while the width
$\Gamma_{tot}$ may be affected by density.

\begin{figure}[htb]
\centerline{\epsfig{file=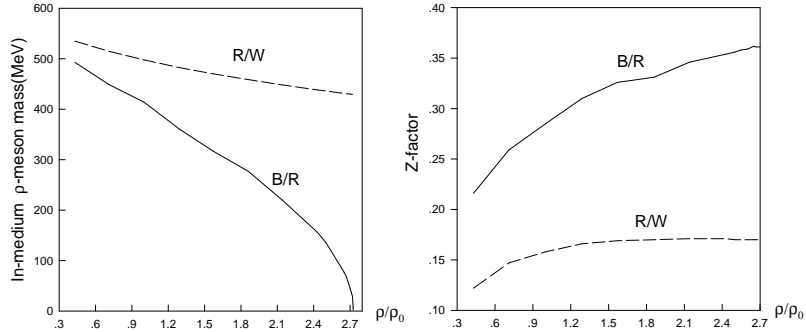,width=11cm}}
\vskip -1.4cm
\caption{\small In-medium $\rho$-meson mass and $Z$ factor
in B/R (solid line) and R/W (dashed line) theories
          for $\Gamma_{tot}=0$.}\label{fig456}
\end{figure}

\vskip 1cm
$\bullet$ \underline{Case of $\Gamma_{tot}=0$}
\vskip 0.3cm

In contrast to the R/W approach, this is a situation which is
actually realizable as density approaches the chiral transition
density $\rho_c$ since the phase space for $\rho$ decay goes to zero
at that density. Let us consider what happens at normal nuclear
density $(\rho_0)$ in the limit of zero width. The solutions are
\be
q_0^-=406.7 ~{\rm MeV},~q_0^+=873.9 ~{\rm MeV}
\ee
with the corresponding $Z$-factors
\be
Z(q_0^-)=0.285,~Z(q_0^+)=0.714
\ee
For $\rho=2.5\rho_0$, we obtain
\be
q_0^-=136 ~{\rm MeV},~q_0^+=956 ~{\rm MeV}
\ee
and
\be
Z(q_0^-)=0.356,~Z(q_0^+)=0.646.
\ee
Since the width should vanish near the critical density $\rho_c$,
the dispersion formula with zero width should approach the correct
one near it. Figure \ref{fig456} shows indeed that
$m_\rho^*\rightarrow 0$ as $\rho\rightarrow ~2.75\rho_0$ as found in
\cite{rho}.

$\bullet$ \underline{Case of $\Gamma_{tot}=\Gamma_0=120$ MeV}
\vskip 0.3cm

The results at normal nuclear density are
\ba
q_0&=&423 ~{\rm MeV},~567 ~{\rm MeV},~878 ~{\rm MeV}\no
Z(423)&=&0.34,~Z(567)=-0.0799,~Z(878)=0.73.
\ea
For $\rho=2.5\rho_0$, we get
\ba
q_0&=&146 ~{\rm MeV},~577 ~{\rm MeV},~951 ~{\rm MeV}\no
Z(146)&=&0.369,~Z(567)=-0.027,~Z(951)=0.657
\ea
We compare in fig.\ref{fig456} the in-medium
$\rho$-meson mass and $Z$-factors in B/R and R/W.

\begin{figure}[htb]
\centerline{\epsfig{file=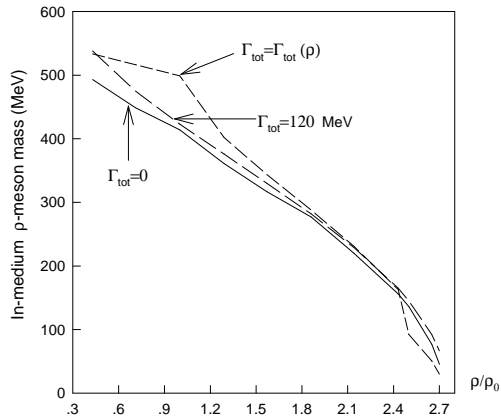,width=8cm}}
\vskip -5.9cm
\caption{\small In-medium $\rho$-meson mass in B/R theory.
        $\Gamma_{tot}=\Gamma_{tot}(\rho)$ corresponds to  the result of
 changing the value of $\Gamma_{tot}$ from $260$ MeV to $30$ MeV 
as density increases.}
\label{fig6}
\end{figure}

\subsection{The $m_\rho^*$ as an order parameter}
\indent\indent
In \cite{rho,adami}, an argument was given that the in-medium mass
of the $\rho$-meson can be taken, roughly, as an order parameter
for the chiral phase transition. Figure \ref{fig6} shows that our
model described above predicts $m_\rho^*$ dropping roughly linearly
in density. This is consistent with the behavior of the quark
condensate in medium,
\be
\frac{\langle\bar qq\rangle^*}{\langle\bar q q\rangle}
\approx 1-\frac{\sigma_N\rho_N}{f_\pi^2m_\pi^2}\label{condensate}
\ee
where the star denotes finite density (or temperature) and $\rho_N$
the nuclear (vector) density. Indeed we would find from
eq.(\ref{condensate}) roughly the same $\rho_c$ as in
fig.\ref{fig456} for $m_\rho^*\sim 0$ by setting $\langle\bar
qq\rangle^*\sim 0$. Thus our model has the quark condensate, on the
average, dropping roughly linearly with density $\rho$.

\section{The $\omega$-meson in nuclear matter}
\indent\indent
In this section,
we apply the same two-level model to
the $\omega$-meson channel. We shall consider both R/W and B/R approaches.
\subsection{The R/W approach}
\indent\indent
For this calculation,
all we have to do is to  replace $f_{\rho N^* N}(m_\rho)$ by
$f_{\omega N^* N}(m_\omega)$ and $m_\rho$ by $m_\omega$
in eq.(\ref{RWself}).
A priori, we do not know how to relate $f_{\omega N^*
N}$ to $f_{\rho N^*N}$. Assuming a generalized VDM would give the
relation $f_{\omega N* N}=3 f_{\rho N* N}$ but there is no
reason, theoretical or empirical, to believe that such a relation
should be reliable. We shall instead resort to the empirical result of Friman
et al~\cite{friman}. From their fig. 4, we find
\be
f_{\omega N^* N}^2\approx 4.4*f_{\rho N^* N}^2\label{factor4.4}
\ee
\vskip 0.3cm
$\bullet$ {\underline{Case of $\Gamma_{tot}=0$}}
\vskip 0.3cm

\begin{figure}[htb]
\vskip -2cm
\centerline{\epsfig{file=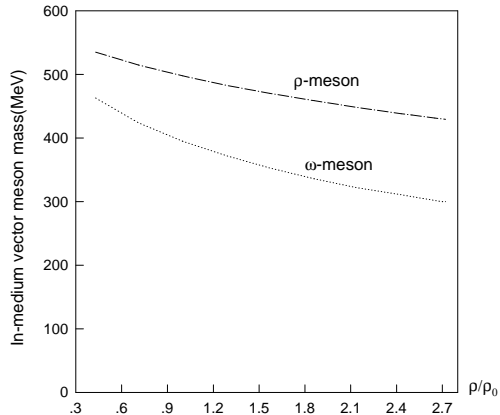,width=8cm}}
\vskip -3.2cm
\caption{\small Comparison of the in-medium ($\rho$ and $\omega$)
vector-meson masses in R/W theory for $\Gamma_{tot}=0$.}\label{mro}
\end{figure}

At normal nuclear matter density, the dispersion formula (corresponding to
eq.(\ref{dis}) for the $\rho$ meson) is
\be
x=1+0.863\frac{x}{x-0.55}
\ee
where $x=q_0^2/m_\omega^2$. The solutions are
\be
q_0^-\simeq 395~{\rm MeV}, ~q_0^+\simeq 1149~{\rm MeV}
\ee
with the corresponding $Z$ factors
\be
Z(q_0^-)\simeq 0.155,~Z(q_0^+)\simeq 0.845.
\ee
The behavior of the $\omega$ mass is compared with that of the
$\rho$ mass in fig.\ref{mro}. Note that the stronger coupling makes the
$\omega$ mass fall faster than the $\rho$ mass.

\vskip 0.3cm
$\bullet$ {\underline{Case of $\Gamma_0=120$ MeV}}
\vskip 0.3cm

In this case, we find
\be
q_0^1=403~{\rm MeV},~q_0^2=576~{\rm MeV}, ~q_0^3=1145~{\rm MeV}
\ee
and
\be
Z(q_0^1)=0.174,~Z(q_0^2)=-0.0293, ~Z(q_0^3)=0.856
\ee
at normal nuclear matter density.
For comparison, we quote the values of Friman et al~\cite{friman}:
\be
q_0^-\simeq 328~{\rm MeV}, ~q_0^+\simeq 1384~{\rm MeV}
\ee
and
\be
Z(q_0^-)\simeq 0.125
\ee
\subsection{The  B/R approach}
\indent\indent
Even if we can extract the $\omega N^* N$ coupling constant from experiments
at zero density,
there is no reason to expect that that constant will remain unchanged in
medium. Indeed we have reasons to believe that the ratio ${\cal R}\equiv
(f_{\omega N* N}/f_{\rho N* N})^2$ will decrease as density increases. 
For this reason, we shall consider two cases: (1) a density-independent 
constant; (2) a density-dependent constant.
\subsubsection{With density-independent $f_{\omega N^* N}$}
\begin{figure}[htb]
\centerline{\epsfig{file=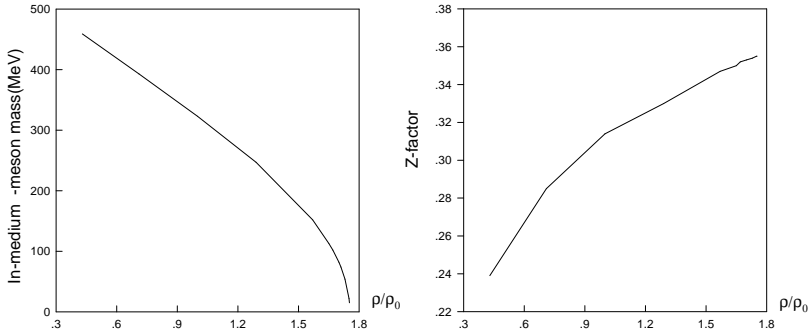,width=11cm}}
\vskip -1.4cm
\caption{\small In-medium $\omega$-meson mass and $Z$ factor
in B/R theory for
        $\Gamma_{tot}=0$ for ${\cal R}$=1.6 independent
of density.}\label{bromz16}
\end{figure}

\indent\indent
With B/R scaling, the factor 4.4 determined empirically
in (\ref{factor4.4}) for matter-free space turns out to give an
unreasonably low critical density ($\rho_c\sim 0.7\rho_0$)
 at which the collective $\omega$
mass vanishes. While the $\omega$ mass is expected to drop faster than
the $\rho$ mass as explained below, it does not seem reasonable that the
$\omega$ mass vanish much before the $\rho$ mass does.  To
see what happens if one takes a constant coupling constant somewhat
larger than the $\rho N^*N$ coupling, we take for illustration 
\be
f_{\omega N^* N}^2\approx 1.6*f_{\rho N^* N}^2\label{factor1.6}.
\ee
We have no particular reason to take this number but it gives a
qualitative idea as to how things go.
The results are summarized in fig.\ref{bromz16} 
 for the case with $\Gamma_{tot}=0$. Since
$|f_{\omega N^* N}|>|f_{\rho N^* N}|$, the $\omega$ mass drops
to zero faster than the $\rho$ mass: 
$m_\omega^*\rightarrow 0$ for $\rho\rightarrow
\rho_c\approx 1.7\rho_0$. 
We shall suggest below that this feature of the $\omega$ properties
may be interpreted in terms of an
``induced symmetry breaking" (ISB) in the vector channel.

\subsubsection{Density-dependent $f_{\omega N^* N}$}
\indent\indent
The fact that empirically ${\cal R}\approx 4$~\cite{friman} at zero density
indicates that the vector dominance model (VDM) -- which would give 9 --
fails. This is not surprising: there is no reason to expect that
the VDM should work in the baryon sector, particularly where baryon resonances
are involved. On the other hand, as density approaches the chiral phase 
transition point, we would expect that the system becomes a Fermi liquid of
quasiquarks~\cite{rho}
to which the vector degrees of freedom corresponding to the $\rho$
and the $\omega$ would couple in an $U(2)$ symmetric way. This would mean that
the ratio ${\cal R}$ would go to 1 as $\rho\rightarrow \rho_c$.

\begin{figure}[ht]
\centerline{\epsfig{file=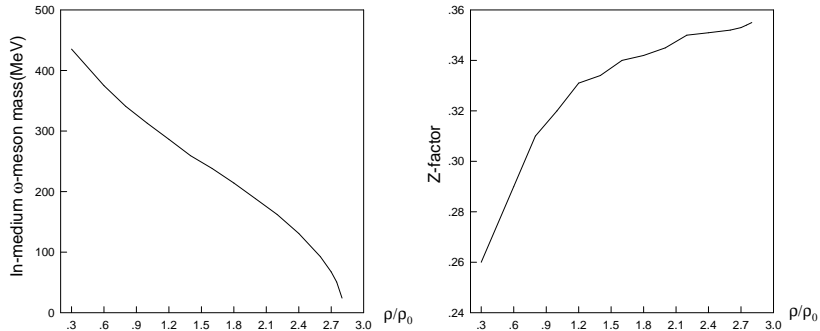,width=11cm}}
\vskip -1.7cm
\caption{\small In-medium $\omega$-meson mass and $Z$ factor 
        in B/R theory for
        $\Gamma_{tot}=0$ with the density dependence of $f_{\omega N^*N}^2$ 
given by eq.(\ref{pa1}).}\label{bromz3}
\end{figure}

Here we shall implement this possibility in the dispersion formula for 
in-medium $\omega$'s assuming that 
the constant $f_{\rho N^*N}$ depends little on density. It turns out that
the simplest possible linear interpolation between $\rho=0$ and $\rho=\rho_c$
gives  too rapid a decrease of the $\omega$ mass, vanishing at much too low a
density than that of the $\rho$. Different parameterizations have been tried
and we report two of them which appear to be reasonable. One is 
the following log-type parametrization
\be
f_{\omega N^* N}^2=f_{\rho N^* N}^2\frac{4.4}
{1+3.4\log(1+1.72(c/2.8))}\label{pa1}
\ee
where $c$ is defined by $\rho=c\rho_0$. 
In this parametrizaton, $\rho_c$ is chosen as
 $\rho_c=2.8\rho_0$.
The results are given in fig.\ref{bromz3}.

\begin{figure}[ht]
\centerline{\epsfig{file=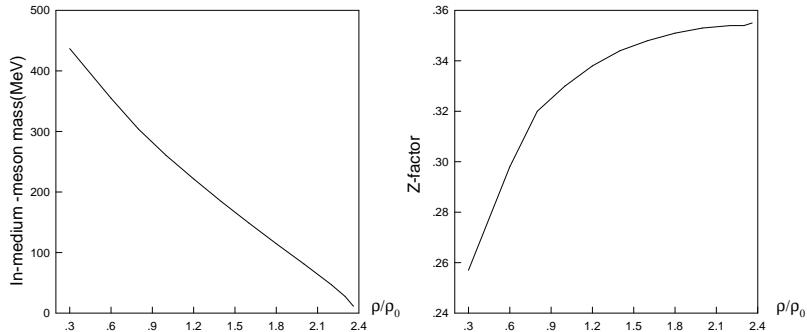,width=11cm}}
\vskip -1.7cm
\caption{\small In-medium $\omega$-meson mass and $Z$ factor
in B/R theory for
        $\Gamma_{tot}=0$ with  the density dependence of $f_{\omega N^*N}^2$ 
given by eq.(\ref{pa2}).}\label{bromzf}
\end{figure}

As an alternative parametrization, we take
\be
f_{\omega N^* N}^2=f_{\rho N^* N}^2
(4.4-3.4(c/2.8)^{\frac{1}{3}}).\label{pa2}
\ee
Figures \ref{bromz3} and \ref{bromzf} show that 
the two parameterizations (\ref{pa1}) and (\ref{pa2}) give qualitatively
the same results\footnote{The density at which the $\omega$ mass goes to 
zero differs slightly but we do not believe that any 
importance can be attached to this difference.}. 
Because of the initially stronger coupling constant, the $\omega$ falls
faster than the $\rho$ at the beginning, flattens in the middle
and then approaches zero at the critical density together with the $\rho$.

\subsection{The dropping $\omega$ mass and a high-density ``ISB phase"}
\indent\indent
The stronger $\omega N^* N$ coupling relative to the $\rho N^* N$
coupling leads naturally to the prediction in the present model
that in medium the $\omega$ mass would fall faster than the $\rho$
mass as density increases. This is expected in both R/W and B/R 
approaches. In B/R, however, this leads to the additional
prediction that the $\omega$ mass would go to zero (in the chiral
limit) either before or at $\rho_c$ for the $\rho$ meson depending on
whether the ratio ${\cal R}$ remains constant (of density) or
goes to 1 at $\rho=\rho_c$. There is no theoretical reason known to
favor one scenario over the other. However that the $\omega$ mass falls faster
than the $\rho$ mass -- which is essentially dictated in the present 
formalism by the fact that ${\cal R} >1$ -- is consistent with the phase
structure of dense matter previously arrived at in
quark language by Langfeld et al~\cite{langfeld1,langfeld2}.

Briefly the scenario given by \cite{langfeld1,langfeld2} is as
follows. If quark-quark interactions have a strength in vector
channel comparable to what is found in one-gluon exchanges, then an
induced Lorentz symmetry breaking could take place at a critical
chemical potential $\mu_c$ at which chiral symmetry would be
restored, i.e., $\langle\bar{q}q\rangle=0$, and the baryon density
$\langle\bar{B}\gamma_0B\rangle$ would have a {\it discontinuous
increase} as $\mu$ exceeds $\mu_c$ indicating a first-order transition.
The consequence is that a low-energy
collective state carrying the quantum number of an $\omega$ meson
should emerge at $\mu_c$ as a pseudo-Goldstone vector boson.

Our proposal is that the collective $N^*$-hole excitation of the
$\omega$ quantum number built in our schematic degenerate model be
identified with the low-mass ISB state described in quark language
in refs.\cite{langfeld1,langfeld2}. In this ``dual" description,
as the inverse $\omega$ propagator vanishes at the point where 
the mass of the lower $\omega$ branch goes to zero, the $\omega$ field
develops an ``induced VEV" $\delta \langle \omega_0\rangle_{\rho_c}$ so that
there would be a discontinuity at $\rho_c$ of the $\langle \omega_0 \rangle$.
Translated into the baryon (or quark) density, this means that there will
be a jump in density at the critical chemical potential $\mu_c$ if one looks
at the density vs. $\mu$. One could think of this as a chiral symmetry
restoration in dense matter in a way analogous to what was obtained in
ERGF (exact renormalization group flow) by Berges et al.~\cite{berges}.
A more appealing way of viewing the present scenario is that 
it provides a hadronic counterpart of the quark-model
scenario of Langfeld et al.\cite{langfeld1,langfeld2}. It is amusing to
note that the ISB phenomenon in the quark sector is encoded in the
empirical fact in the hadronic sector that $|f_{\omega N^*N}| > |f_{\rho
N^*N}|$ for $\rho<\rho_c$.

\section{Conclusion}
\indent\indent
We have constructed a schematic model of the Rapp/Wambach theory,
emphasizing the role of the $[N^* (1520) N^{-1}]$ isobar-hole state.
This model turns out to be essentially the same as the degenerate
schematic model of Brown~\cite{gerry} but for the $q_0$-dependence
of the coupling of $\rho$ meson to $N^*N$. In fact when this
coupling is cancelled by introducing $m_\rho^*$ as the mass scaling
factor of the Lagrangian, the model becomes precisely that of
Brown~\cite{gerry}. We show that this latter model gives the same
results as Brown/Rho scaling in the limit of $\rho\rightarrow
\rho_c$, where $\rho_c$ is the chiral restoration density and propose
to use it as an interpolation formula between R/W and B/R scaling.
This then provides a possible mechanism to arrive at B/R scaling
from the hadronic side, that is, in the bottom-up way.

New results from the TCP now in service with the CERES collaboration
should pin down the strength at the $\rho$-meson poles, or at least
an average of this strength over the various densities encountered
in this experiment. These will confirm or infirm our scenario:
Since the expected strength entering into the
dilepton rate is obtained by
$A(\omega)=\frac{Z(\omega)}{2\omega}$,
lower-energy component(s) will be progressively more enhanced at
higher densities. Furthermore, given temperature of $T\sim 150$ MeV, there
will be larger Boltzmann factors at lower energies, so the net result will
be that more leptons will come out of the lower-energy state(s).

We have also suggested that the $\omega$ mass should fall faster
than the $\rho$ mass until one approaches the chiral phase transition
and that the collective $N^*$-hole excitation of the $\omega$
quantum number in the schematic degenerate model is the hadronic
("Cheshire-cat") description of the pseudo-Goldstone vector
boson generated by an induced symmetry breaking of Lorentz symmetry in
dense medium.
\subsection*{Acknowledgments}
\indent\indent
We are grateful for helpful discussions with Kurt Langfeld and Hyun Kyu Lee.
One of us (YK) is indebted to Hyun Kyu Lee for support and encouragement.
The work of YK was partially supported by KOSEF (Grant No. 985-0200-001-2) 
and the U.S. Department of Energy under Grant No. DE--FG02--88ER40388
and that of RR and GEB by
the U.S. Department of Energy under Grant No. DE--FG02--88ER40388.

\end{document}